\documentstyle[psfig,conf_iap,10pt]{article}
\begin{document}
\newcommand{\lya}{Ly$\alpha$\ }
\newcommand{\etal}{et al.\ }
\heading{%
%
New Statistical Measures of the Ly$\alpha$ Forest Spectra for
Accurate Comparison to Theoretical Models $^1$

%
} 
\par\medskip\noindent
\author{%
Jordi Miralda-Escud\'e $^{2,8}$,
Michael Rauch$^{3,9}$, 
Wallace L.W. Sargent$^3$,
Tom A. Barlow$^3$,
David H. Weinberg$^4$, Lars Hernquist$^{5,10}$,
Neal Katz$^6$, Renyue Cen$^7$,  
Jeremiah P. Ostriker$^7$

}
\address{%
The observations were made at the W.M. Keck Observatory
which is operated as a scientific partnership between the California
Institute of Technology and the University of California; it was made
possible by the generous support of the W.M. Keck Foundation.
}
\address{%
Department of Physics and Astronomy,
University of Pennsylvania, Philadelphia, PA 19104
}
\address{%
Astronomy Department, California Institute of Technology,
Pasadena, CA 91125, USA
}
\address{%
Department of Astronomy, The Ohio State University, Columbus, OH 43210
}
\address{%
Lick Observatory, University of California, Santa Cruz, CA 95064
}
\address{%
Department of Physics and Astronomy, University of Massachusetts, Amherst, MA, 98195
}
\address{%
Princeton University Observatory, Princeton, NJ 08544
}
\address{%
Alfred P. Sloan Fellow
}
\address{%
Hubble Fellow
}
\address{%
Presidential Faculty Fellow
}

\begin{abstract}
We propose a new method of analysis for the \lya forest, namely to
measure the 1-point and 2-point joint probability distribution of the
transmitted flux. The results for a sample of seven observed quasars
and from two simulations of structure formation are shown and compared.
Statistically significant differences in the 2-point function between
the results of the numerical simulations and the observations are
easily found. The analysis we suggest is very simple to apply to
observed data sets, and we discuss its
superiority over the traditional Voigt-profile fitting algorithms for
accurate comparison to the predictions of theoretical models.

\end{abstract}
\section{ Introduction}

  It has recently been shown that the \lya forest is a natural
prediction of large-scale-structure models that have been adjusted
to fit independent observations of the spatial distribution of
galaxies (e.g., \cite{cen94}, \cite{hkwm96}, \cite{mcor96},
\cite{zhang97}). Figure 1 shows 10 spectra from
the simulation L10 in \cite{mcor96} at $z=3$
(left column), and 10
small pieces of the spectrum of the quasar Q1422+230 of the same length,
near $z=3$ (right column;
a full analysis of these observations will be presented elsewhere).
Simple examination by eye does not reveal obvious
differences. Accurate statistical methods must be used to measure
any subtle differences between the observations and the predictions
of different theoretical models.

\begin{figure}
\vskip -1.cm
\centerline{\vbox{
\psfig{figure=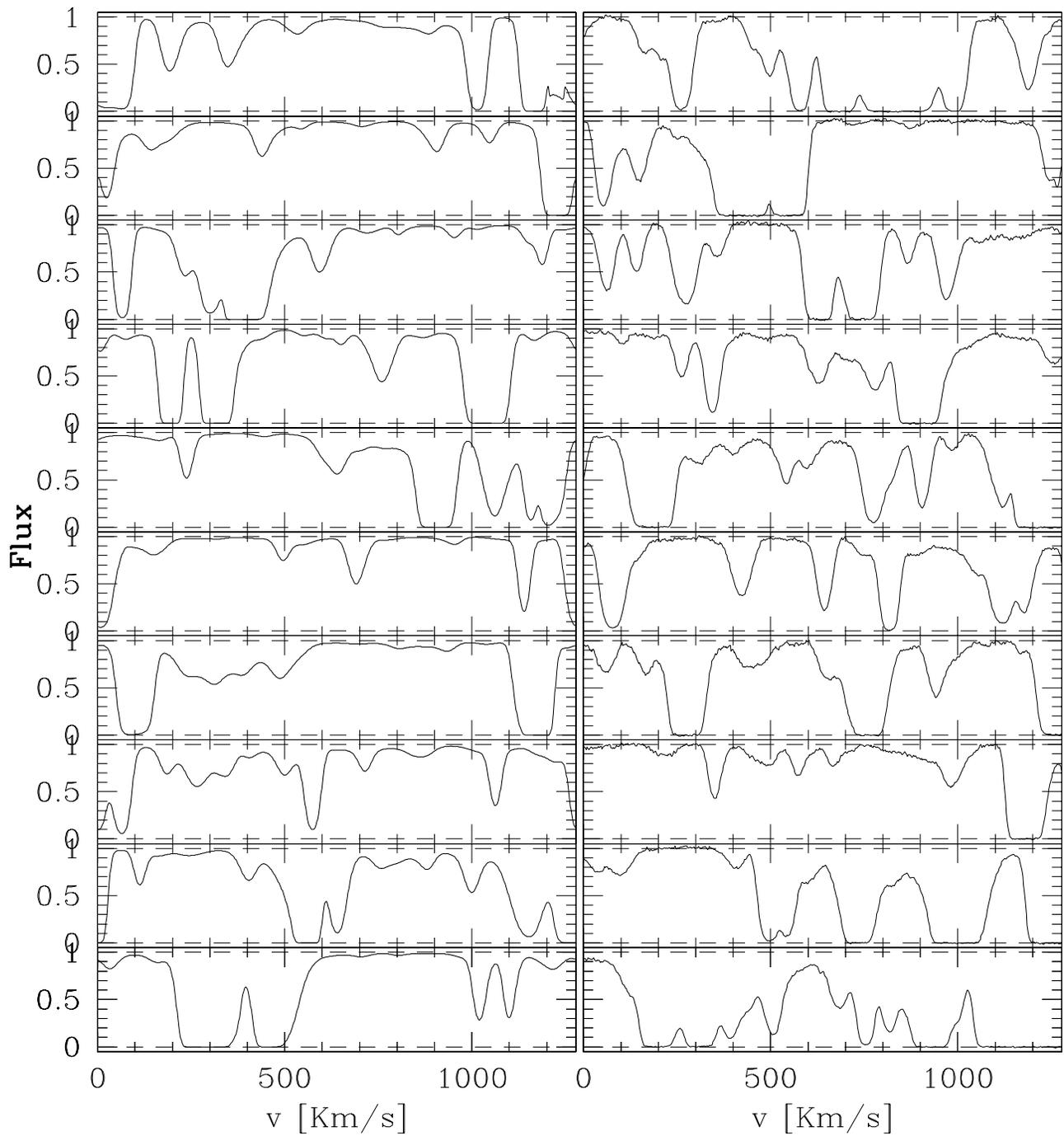,height=20.cm}
}}
\caption[]{{\it Left column:} Ten random spectra of a cosmological
simulation of the \lya forest. {\it Right column:} Ten randomly chosen
small pieces of the observed spectrum of Q1422+230 }
\end{figure}

  The \lya forest arises from density fluctuations in the intergalactic
medium caused by the collapse of structure in a network of filaments and
sheets. Thus, the observed spectrum is
a one-dimensional function depending on the density, temperature and
velocity of the gas in every point along the line-of-sight, and is
not caused by individual, isolated clouds producing absorption lines.
Traditionally, the \lya forest has been analyzed by selecting
absorption lines and fitting them to Voigt profiles, using the
operation of ``deblending'' into multiple lines whenever an
absorption feature is not well fitted by a single Voigt profile.
This method of analysis is inadequate in all cases where true
fluctuations in the transmitted flux are detected over a large fraction
of the spectrum. There are several reasons for this
inadequacy: the deblending operation is extremely complex and
involves a large number of arbitrary parameters needed for
selecting fitting regions, deciding the number of superposed
absorption lines needed for every fit, and choosing initial guesses
for the fit which may determine which local minimum of the $\chi^2$
function in the parameter space is chosen as a solution. These
problems become more severe as the quality of the data improves
owing to the increased number of components needed for the fits,
and the results obtained (such as the distribution of column densities
and Doppler parameters, and the line correlation function) will in
general not converge to a fixed answer as the signal-to-noise in a
spectrum is increased, and will depend on details of the algorithm.
In addition, there is no possible physical interpretation of
quantities such as the line correlation function, because the
fitted lines with superposed profiles do not actually correspond to
any physical objects that exist, as shown by the simulations of the
\lya forest.

\section{Statistical Distributions of the Transmitted Flux}
The most simple method of analysis must be based on direct use of the
observed quantity, the transmitted flux in every pixel.
All the statistical information that can be inferred from observations
is contained in the N-point joint probability distribution of the
transmitted flux. Thus, the obvious quantities to measure first are
the flux probability distribution, and the 2-point joint probability
distribution.

  The flux probability distribution (first discussed by \cite{jo91}
and \cite{webb92}) was obtained from observations of
seven quasars in \cite{rauch97}, and compared to the results of
numerical simulations of two cold dark matter models: a model with a
cosmological constant using an Eulerian simulation ($\Lambda$CDM), and
a model with
$\Omega=1$ run with a Tree-SPH code (SCDM). For details on the models,
see \cite{rauch97} and references therein. Figure 2 shows the
flux distribution at $z=3$; for the $\Lambda$CDM model, the
distribution is shown as obtained directly from the simulated spectra
(labeled $\Lambda$CDM, raw), and after corrections to simulate the
fitting of the continuum, instrumental resolution and noise in the
observations. These corrections are fully described in \cite{rauch97};
the figure shows that the modifications introduced by the corrections
are not too large. The flux distribution is fairly well fit by the
two models. Only one parameter in the simulations has been adjusted to
the observation, determining the mean flux decrement. The parameter
is proportional to the square of the baryon density divided by the
intensity of the ionizing background, so these quantities can be
constrained (see \cite{rauch97}).

\begin{figure}
\centerline{\vbox{
\psfig{figure=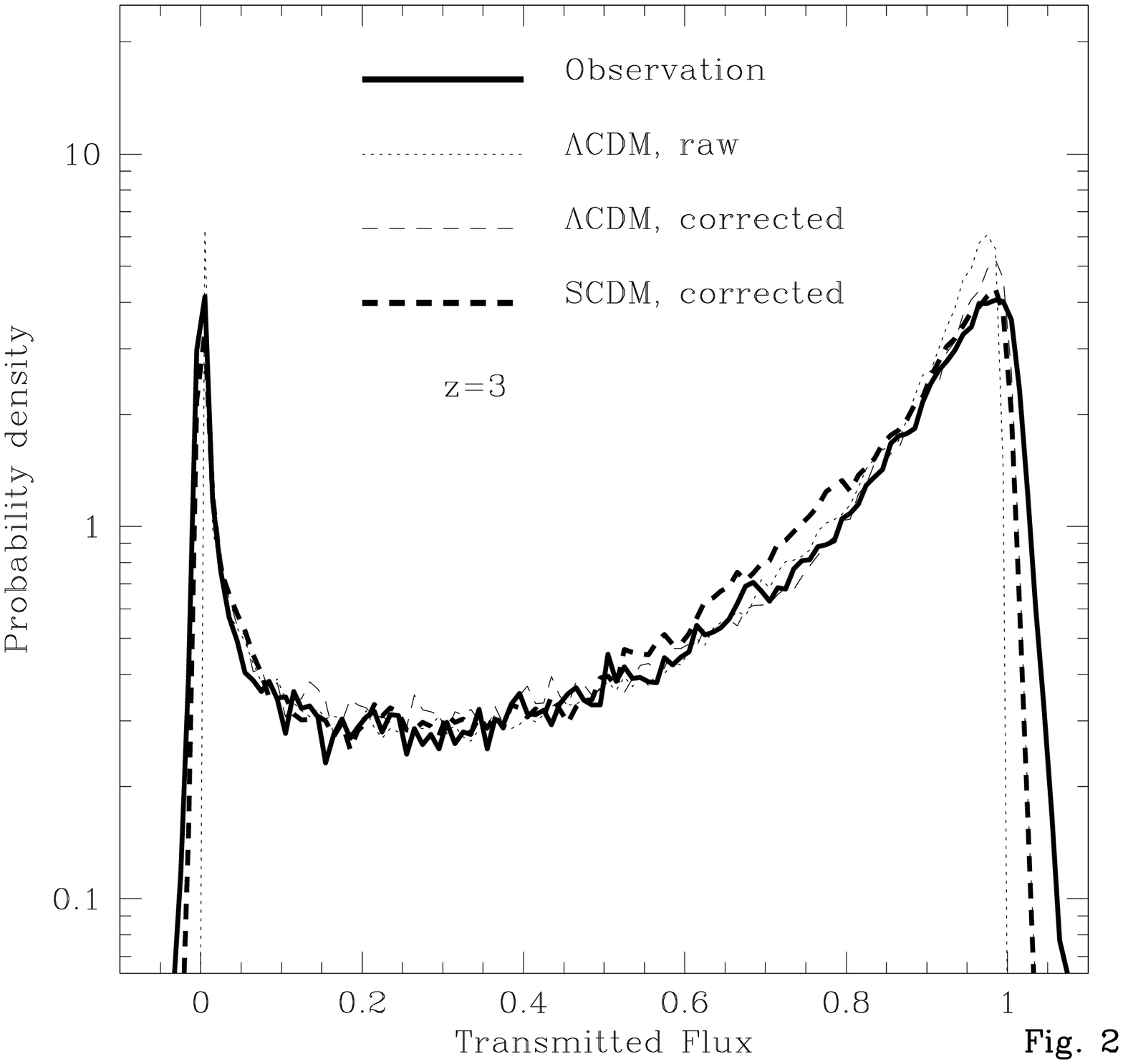,height=10.cm}
}}
\end{figure}

  The two-point function $P_2(F_1, F_2, \Delta v)$ is the probability
that two randomly chosen pixels separated in the spectrum by a
velocity interval $\Delta v$ will have transmitted flux $F_1$ and
$F_2$. A convenient way to visualize this function is by defining
moments over $F_2$:
\begin{equation}
S_m(F, \Delta v) \equiv \int_0^1 dF_2\, P_2(F, F_2, \Delta v)\, (F-F_2)^m
 ~. \end{equation}
Here, we shall show results only for $m=1$, which is the mean flux
difference. The results for $S_1$ are shown in Figure 3 as a function
of $\Delta v$, when we average $S_1$ over the intervals of the flux $F$
indicated in the figure. The shape of these curves yields information
about the mean shape of the absorption features. For example, for
$0< F < 0.1$, the first pixel is always near the saturated part of an
absorption profile, and the shape of the curve depends on the mean
profile of saturated absorbers, while the $0.6 < F < 1$ curve is more
sensitive to weak absorbers. Figure 3 shows that in the $\Lambda$CDM
simulation the strong absorbers are narrower than in the observations
(in the traditional language, their Doppler parameters are too small or,
at large $\Delta v$, the line correlation is too weak), while weak
absorbers are approximately as observed. On the other hand, the SCDM
simulation has weak absorbers that are too broad, and strong absorbers
with approximately the observed widths.

\begin{figure}
\centerline{\vbox{
\psfig{figure=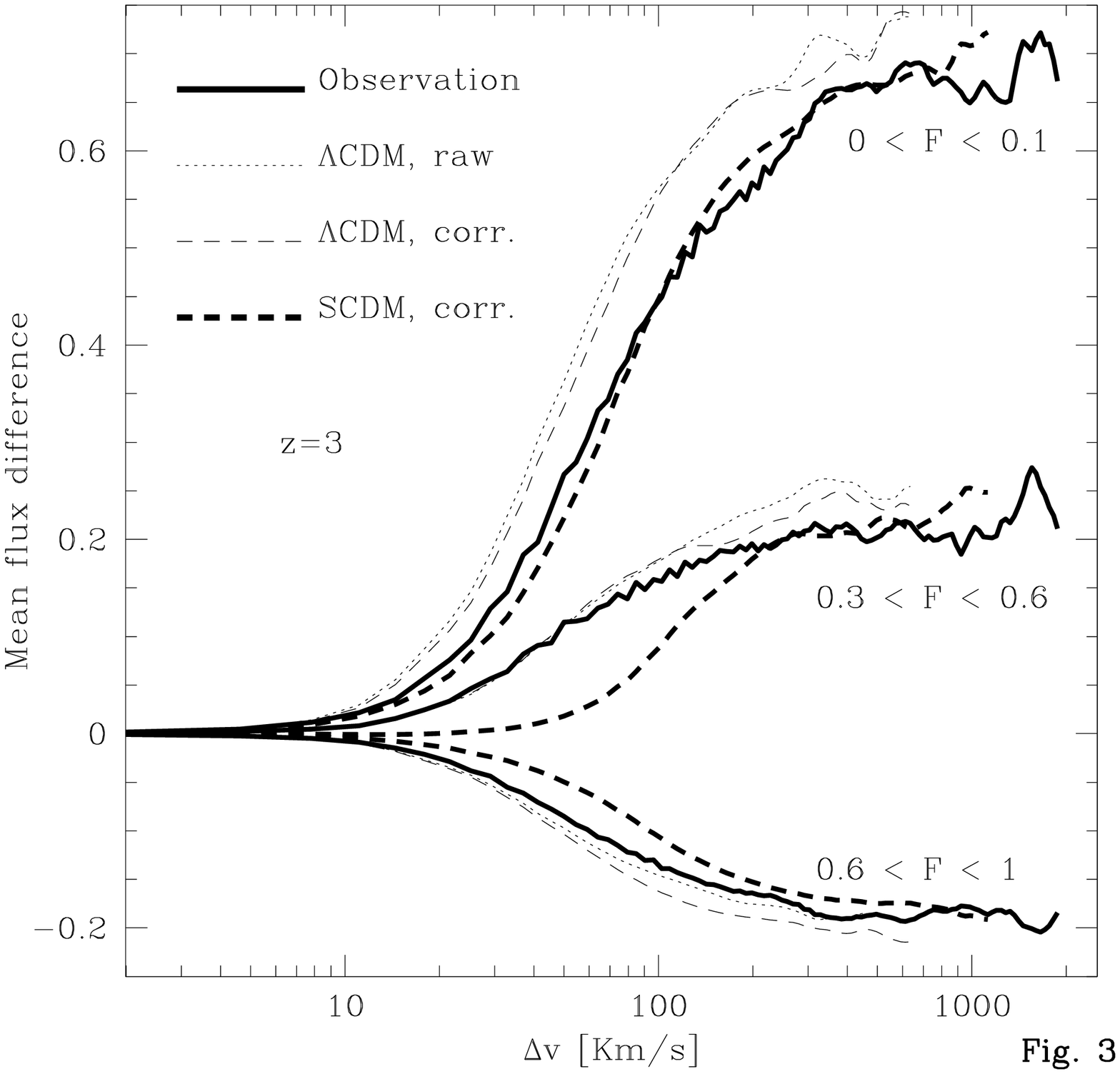,height=10.cm}
}}
\end{figure}

  It is not at all clear at present if these differences
represent a true failure of the cosmological models assumed, or arise
as a result of errors in the simulations due to the limited dynamic
range and the effect of physical processes that are not included.
Future work will need to concentrate on evaluating errors in the
observations and theoretical predictions, which we discuss briefly
in the next section.

\section{Discussion}

  The largest source of error in the observations is due to the small
sample of observed quasars (only 3 quasars contribute significantly to
the $z=3$ measurements in Figs. 2 and 3). We have estimated the error
from the difference in the results when we divide our sample into two
parts. While the largest difference in the flux distribution function
(Fig. 2) with the SCDM model, in the interval $0.6 < F < 0.8$, is only
of marginal significance, the observational error of the curves in Fig.
3 is much smaller than the difference with the two models mentioned
above.  This proves that both of these
models, which were found to be in rough agreement with observations
in preliminary comparisons (\cite{hkwm96}, \cite{mcor96}), show clear
(even though not very large) differences with
observations once the comparison is made accurately, using robust
statistical methods. This will be presented in much more detail in
a paper in preparation on the 2-point function.

  This proves only that the observational results are clearly different
from our calculation of these two models, using numerical simulations.
The predictions from
cosmological simulations are subject to two types of theoretical
errors. First is the error due to the limited dynamic range of the
simulations, arising both from the finite numerical resolution and from
the small size of the simulated boxes. As an example of the errors that
are introduced, the power spectrum of the models is truncated at the
scale of the simulated box, and the absence of the large-scale power
reduces the large-scale velocities, which may result in reducing the
width of the absorption features. Detailed studies will be needed to
quantify the magnitude of these errors on results such as those
presented here; it is possible that these errors are as large as the
differences between models and observations we have found. We emphasize
that the two models we use were computed with two different
numerical codes, so these errors may be quite different in nature
for the two simulations.

  The second type of theoretical error may be due to a physical
modification of the model. The numerical simulations assume that the
gas is only affected by gravity and the pressure force resulting from
photoionization by a homogeneous radiation background. Among the
modifications that this simplified picture could have in the real
universe are an inhomogeneity in the heating due to photoionization,
and shock heating caused by gas ejection from starburst galaxies or
AGNs.

  The amount of data that can be gathered on the \lya forest is
extremely large, and many statistical properties can be measured in
addition to the functions suggested here. It is therefore possible
that one can learn both about the underlying theory for
the initial density fluctuations that gave rise to the \lya forest
and about other possible effects that may have influenced the
evolution of the intergalactic medium. The use of the
observational data to constrain models clearly requires a
quantification of the errors in the predictions made from simulations,
due to the effects mentioned above. But even before we have more
robust theoretical predictions from different models, the observational
determination of new statistical properties of the \lya forest can
advance independently.





%
\vfill

\begin{iapbib}{99}{
\bibitem{cen94}
Cen, R., Miralda-Escud\'e, J., Ostriker, J. P., Rauch, M., 1994, \apj, 437, L9 
\bibitem{hkwm96}
Hernquist L., Katz N., Weinberg D.H.,  Miralda-Escud\'e J., 1996, \apj, 457, L5
\bibitem{jo91}
Jenkins, E. B., \& Ostriker, J. P. 1991, \apj, 376, 33
\bibitem{mcor96}
Miralda-Escud\'e J., Cen R., Ostriker J. P., Rauch M., 1996, \apj, 471, 582
\bibitem{rauch97}
Rauch, M., Miralda-Escud\'e, J., Sargent, W. L. W., Barlow, T. A.,
Weinberg, D. H., Hernquist, L., Katz, N., Cen, R., Ostriker, J. P.,
1997, \apj, submitted (astro-ph/9612245)
\bibitem{webb92} Webb, J. K., Barcons, X., Carswell, R. F., Parnell, H. C.,
1992, MNRAS, 255, 319
\bibitem{zhang97} Zhang, Y, Anninos, P., Norman, M. L., \& Meiksin, A. 1997,
\apj, 485, 496
%
}
\end{iapbib}
\end{document}